\documentclass{elsart}
\usepackage{ifpdf}
\usepackage{natbib,amssymb,amsmath}
\ifpdf
\usepackage[%
  pdfstartview=FitH,%
  bookmarks=true,%
  bookmarksopen=true,%
  breaklinks=true,%
  colorlinks=true,%
  linkcolor=blue,anchorcolor=blue,%
  citecolor=blue,filecolor=blue,%
  menucolor=blue,pagecolor=blue,%
  urlcolor=blue]{hyperref}
\else
\usepackage[%
  breaklinks=true,%
  colorlinks=true,%
  linkcolor=blue,anchorcolor=blue,%
  citecolor=blue,filecolor=blue,%
  menucolor=blue,pagecolor=blue,%
  urlcolor=blue]{hyperref}
\fi

\journal{Stud.\ Hist.\ Phil.\ Mod.\ Phys.}

\newcommand{\ket}[1]{\ensuremath{|{#1\rangle}}} 

\newcommand{\braket}[2]{\ensuremath{{\langle #1}|{#2 \rangle}}}
\newcommand{\ketbra}[2]{\ensuremath{|{#1 \rangle}{\langle #2}|}}
\providecommand{\abs}[1]{\left\lvert#1\right\rvert}
\newcommand{\op}[1]{\hat{#1}}

\begin{document}

\begin{frontmatter}

  \title{The quantum-to-classical transition:\\Bohr's doctrine of
    classical concepts,\\emergent classicality, and decoherence}

\author[Physics]{Maximilian Schlosshauer\corauthref{cor1}},
\ead{m.schlosshauer@unimelb.edu.au}
\corauth[cor1]{Corresponding author.}
\author[HPS]{Kristian Camilleri} 

\address[Physics]{School of Physics, University of Melbourne,
  Melbourne, Victoria 3010, Australia}

\address[HPS]{History and Philosophy of Science Department, University
  of Melbourne, Melbourne, Victoria 3010, Australia}

\begin{abstract}
  It is now widely accepted that environmental entanglement and the
  resulting decoherence processes play a crucial role in the
  quantum-to-classical transition and the emergence of
  ``classicality'' from quantum mechanics. To this extent, decoherence
  is often understood as signifying a break with the Copenhagen
  interpretation, and in particular with Bohr's view of the
  indispensability of classical concepts. This paper analyzes the
  relationship between Bohr's understanding of the quantum--classical
  divide and his doctrine of classical concepts and the
  decoherence-based program of emergent classicality. By drawing on
  Howard's reconstruction of Bohr's doctrine of classical concepts,
  and by paying careful attention to a hitherto overlooked
  disagreement between Heisenberg and Bohr in the 1930s about the
  placement of the quantum--classical ``cut,'' we show that Bohr's
  view of the quantum--classical divide can be physically justified by
  appealing to decoherence. We also discuss early anticipations of the
  role of the environment in the quantum--classical problem in
  Heisenberg's writings. Finally, we distinguish four different
  formulations of the doctrine of classical concepts in an effort to
  present a more nuanced assessment of the relationship between Bohr's
  views and decoherence that challenges oversimplified statements
  frequently found in the literature.
\end{abstract}


\end{frontmatter}

\section{Introduction}

Bohr's interpretation of quantum mechanics, first sketched in his Como
lecture of 1927 \citep{Bohr:1927:zb} and subsequently developed in his
papers in the late 1920s and 1930s, is characterized by two main
features. The first feature is the assumption that \emph{classical
  concepts} are indispensable for describing the results of
experiments involving quantum phenomena, in spite of their limited
applicability. The second feature is the concept of
\emph{complementarity}, which describes the need to use mutually
exclusive experimental arrangements in the use of classical concepts
such as position and momentum \citep{Camilleri:2007:um}. While the
notion of complementarity is by far the more prominent of the two, as
\citet[p.~202]{Howard:1994:lm} explains, ``the doctrine of classical
concepts turns out to be more fundamental to Bohr's philosophy of
physics than are better-known doctrines, like complementarity.'' In
perhaps his clearest expression of the doctrine, Bohr argued:

\begin{quote}
  It is decisive to recognize that, \emph{however far the phenomena
    transcend the scope of classical physical explanation, the account
    of all evidence must be expressed in classical terms.}  The
  argument is simply that by the word ``experiment'' we refer to a
  situation where we can tell others what we have done and what we
  have learned and that, therefore, the account of the experimental
  arrangement and of the results of the observations must be expressed
  in unambiguous language with suitable application of the terminology
  of classical physics \citep[p.~209]{Bohr:1949:mz}.
\end{quote}
Scholars have long pondered over precisely what Bohr meant by
``classical concepts'' or why he felt they should play such a primary
role in quantum physics. Bohr's doctrine was a controversial one and
remains the subject of much debate and disagreement to this day. Bohr
believed quantum physics to be the universally correct theory, which
thus would in principle---i.e., given an appropriate experimental
arrangement---have to also apply to the description of macroscopic
measurement apparatuses and observers. However, as the passage above
indicates, Bohr felt that the experimental setup must be described in
terms of classical physics, if it is to serve as a measuring
instrument at all. One is left with the impression from Bohr's
writings that the quantum--classical divide is a necessary part of the
epistemological structure of quantum mechanics. This view finds
expression in the works of physicists like Heisenberg and Rosenfeld,
who were deeply influenced by Bohr.

We may therefore raise two questions about the views of Bohr and his
contemporaries, such as Heisenberg. First, what exactly was the
meaning, justification, and location of the quantum--classical divide?
Second, how were transitions between the quantum and the classical
realm, i.e., the ``crossings'' of the quantum--classical boundary,
understood and explained---in particular, in what sense and how may
classicality arise from within quantum mechanics?  These questions lie
at the heart of many interpretations of quantum mechanics whose main
goal is to solve the so-called measurement problem---a merely
historically motivated term that should be more appropriately subsumed
under the general heading of the \emph{problem of the
  quantum-to-classical transition}. Remarkably, recent developments in
the study of decoherence \citep[for reviews,
see][]{Joos:2003:jh,Zurek:2002:ii,Schlosshauer:2003:tv,Schlosshauer:2007:un}
have shown that it is possible to address this problem (at least
partially) by realizing that all realistic quantum systems are
inevitably coupled to their environment.  Studies of decoherence have
shed new light on the emergence of classical structures from within
the quantum realm and have led to enormous progress in a quantitative
understanding of the quantum-to-classical transition.

The natural question is then to ask to what extent such a research
program may run counter to Bohr's assumptions of intrinsic,
underivable classical concepts and of different, mutually
contradictory descriptions as embraced by the complementarity
principle. Can fuzzily defined concepts such as measurement,
quantum--classical dualism, and complementarity be reduced to
effective concepts derivable in terms of unitarily evolving wave
functions and the influence of decoherence, and can they thus shown to
be but superfluous semantic or philosophical baggage? Needless to say,
this is an extremely complex question pertaining to many issues of
both theory and interpretation.

Questions concerning the relationship between the decoherence approach
and Bohr's interpretation of quantum mechanics have been raised by a
number of physicists working on decoherence. However, there has been
little serious and nuanced investigation on this matter. For example,
Zeh has contrasted the dynamical approach of decoherence with the
``irrationalism'' of the Copenhagen school
\citep[p.~27]{Joos:2003:jh}. However, one may object that this
characterization perpetuates the familiar myth of the Copenhagen
interpretation, and in particular Bohr's viewpoint, as entailing some
kind of radical subjective idealism. It is our view that a deeper
understanding of the relationship between the views of Bohr and his
followers and the program of decoherence merits more careful
historical and philosophical investigation.

This paper is organized as follows.  Sec.~\ref{sec:quant-class-trans}
prepares the way by sketching the relevant components of the problem
of the quantum-to-classical transition
(Sec.~\ref{sec:probl-quant-class}) and by providing a brief review of
the decoherence program (Sec.~\ref{sec:basics-decoherence}). In
particular, we will discuss to what extent, and in what sense,
decoherence may be said to explain the emergence of classicality
(Sec.~\ref{sec:emerg-class-thro}).

Secs.~\ref{sec:bohr-heis-quant} and \ref{sec:doctr-class-conc} then
examine relationships between the decoherence account of the
quantum-to-classical transition and Bohr's doctrine of classical
concepts, in the following two ways.

First, in Secs.~\ref{sec:bohr-quant-class} and
\ref{sec:bohr-isol-syst}, we will investigate the degree to which
Bohr's understanding of the \emph{quantum--classical divide} may be
recovered as emergent, and how this may challenge or support both
Bohr's interpretation of quantum mechanics and his intuition about
such concepts. Inevitably, given the large degree of scholarly dispute
about the exact meaning of Bohr's writings and his views, this poses a
difficult problem. Rather than take into account different possible
readings of Bohr's philosophy, we will focus specifically on his view
of the quantum--classical divide in the context of the interaction
between object and measuring apparatus, and the way it may differ
crucially from Heisenberg's interpretation. A close examination of
Bohr's and Heisenberg's writings and correspondence in the 1930s
reveals an underlying disagreement on how the dividing line, or
``cut,'' is to be understood
(Sec.~\ref{sec:heisenberg-cut-bohr}). This point has often gone
unnoticed. We can gain a deeper insight into Bohr's position, and why
he disagreed with Heisenberg on the cut, through Howard's
reconstruction of Bohr's doctrine of classical concepts
(Sec.~\ref{sec:howards-reconstr-boh}). This, in turn, sheds new light
on the points of convergence and divergence between Bohr and the
current decoherence-based program of emergent classicality. We will
also show (Sec.~\ref{sec:heis-entangl-extern}) how certain passages of
Heisenberg's writing point out the importance of the openness of
quantum systems in the problem of classicality. 

Second, in Sec.~\ref{sec:doctr-class-conc} we look in more detail at
the \emph{doctrine of classical concepts}, as it was understood by
Bohr and his followers, in particular Heisenberg, Weizs\"acker, and
Rosenfeld. While their views have often been taken as representative
of what is commonly called the ``Copenhagen interpretation,'' we must
be clear that as it is commonly used, this term refers to a range of
different physical and philosophical perspectives which emerged in the
decades following the establishment of quantum mechanics in the late
1920s.\footnote{As Jammer points out in his \emph{Philosophy of
    Quantum Mechanics}, ``the Copenhagen interpretation is not a
  single, clear-cut, unambiguously defined set of ideas but rather a
  common denominator for a variety of related viewpoints. Nor is it
  necessarily linked with a specific philosophical or ideological
  position'' \cite[p.~87]{Jammer:1974:pq}. Indeed the very idea of a
  unitary interpretation only seems to have emerged in the 1950s in
  the context of the challenge of Soviet Marxist critique of quantum
  mechanics, and the defense of Bohr's views, albeit from different
  epistemological standpoints, by Heisenberg and Rosenfeld.}  Indeed,
a close reading reveals that there is not one single version of Bohr's
``doctrine of classical concepts'' which emerges from the writings of
the Copenhagen school (Sec.~\ref{sec:doctr-class-conc-revis}). Here we
will draw a distinction (Sec.~\ref{sec:pragm-vers-categ}) between
those who defended the view that we \emph{must use} classical concepts
in quantum mechanics (Bohr) and those who took a more pragmatic
position, in arguing that it is simply the case that we \emph{do use}
classical concepts (Weizs\"acker). A key to understanding the
relationship between decoherence and the doctrine of classical
concepts is the distinction which can be drawn between an
\emph{epistemological} and \emph{physical} formulation formulation of
the doctrine (Sec.~\ref{sec:epist-vers-phys}). In
Sec.~\ref{sec:decoh-doctr-class} we shall discuss to what extent
decoherence can be regarded as providing a \emph{physical} account of
Bohr's doctrine.

This paper avoids the tendency which has become customary in much of
the literature to compare the newly emerging ideas on the foundations
of quantum mechanics with some reconstructed, or worse still, some
imagined, version of the ``Copenhagen interpretation.'' Instead we
focus on what Bohr had to say, as well as on the writings of
Heisenberg, Rosenfeld, and Weizs\"acker. To what extent their views
are in stark contrast with Bohr's interpretation or when they
represent an approach entirely consistent with his general viewpoint
is particularly relevant in the context of our discussion of
decoherence. Decoherence marks the most successful attempt of
explaining the quantum-to-classical transition wholly within the
framework of quantum mechanics. But this can be seen as part of a
general approach which a number of Bohr's followers---notably
Weizs\"acker and Rosenfeld---began to pursue in the 1960s. Far from
seeing it as an invalidation of Bohr's basic insight, they regarded it
as providing a justification of his views. Of course, whether this was
true to Bohr's original vision remains a point of some
conjecture. However, it would be just as premature and foolish to
declare Bohr's views (an \emph{interpretation} of quantum mechanics)
as invalidated by decoherence (a \emph{consequence} of quantum theory
and a physical process), as to deny that decoherence suggests a
reconsideration of the status of Bohr's concepts. It is our hope that
our investigation will establish a more refined view on the complex
connections between Bohr's views and those of his followers on the one
hand, and the insights gathered from decoherence and the related
decoherence-inspired interpretive programs on the other hand.

\section{\label{sec:quant-class-trans}The quantum-to-classical
  transition and decoherence}

In this section, we shall briefly review the problem of the
quantum-to-classical transition and the implications of decoherence
for this problem. This will allow us to clearly state the scope of
decoherence as required for a careful comparison with Bohr's views on
the quantum-to-classical transition. The basic formalism of quantum
measurement and decoherence will also be relevant to Howard's
reconstruction of Bohr's doctrine of classical concepts discussed in
Sec.~\ref{sec:howards-reconstr-boh} below.

\subsection{\label{sec:probl-quant-class}The problem of the
  quantum-to-classical transition in quantum mechanics}

Broadly speaking, the problem of the quantum-to-classical transition
is concerned with the difficulty of how to reconcile the
quantum-mechanical description of systems by unitarily evolving wave
functions---which may, for example, describe coherent superpositions
of macroscopically distinguishable states---with the fact that objects
in the macroscopic world are perceived to be in well-defined, robust
``classical'' states and not in superpositions thereof. Specifically,
we may distinguish three related but distinct problems
\citep{Schlosshauer:2007:un}:

\begin{enumerate}

\item \emph{Preferred-basis problem.} What singles out the set of
  ``preferred'' physical quantities in nature? For example, why do we
  observe macroscopic systems to be in definite positions rather than
  in superpositions of positions?

\item \emph{Nonobservability of interferences.}  Why is it so
  prohibitively difficult to observe interference effects, in
  particular on macroscopic scales?

\item \emph{Problem of outcomes.}  How can we explain the apparent
  probabilistic selection of definite outcomes in measurements?

\end{enumerate}

A fourth aspect is the apparent insensitivity of macroscopic systems
to measurements: Measurements on closed quantum systems usually alter
the state of the system, posing the question of how states become
``objectified'' in the sense of classical physics.

The problem of the quantum-to-classical transition is often
illustrated in the context of the von Neumann measurement scheme
\citep{Neumann:1932:gq}, which demonstrates how microscopic (and thus
``unproblematic'') superpositions are readily amplified to the
macroscopic realm. Here one considers interactions of a quantum system
$\mathcal{S}$ with an apparatus $\mathcal{A}$ which is \emph{also}
treated quantum-mechanically. Suppose the apparatus measures a set of
states $\{ \ket{s_n} \}$ of the system in the sense that the
system--apparatus interaction is of the form $\ket{s_n}\ket{a_0} \,
\stackrel{t}{\longrightarrow} \, \ket{s_n}\ket{a_n}$, where
$\ket{a_0}$ is the initial ``ready'' state of the apparatus.  The
linearity of the Schr\"odinger equation then implies that the
tensor-product state of an arbitrary initial state $\sum_n c_n
\ket{s_n}$ of the system and the initial apparatus state $\ket{a_0}$
will evolve into a composite entangled state according to
\begin{equation}
  \label{eq:2adgdvaj}
  \left(  \sum_n c_n \ket{s_n} \right)\ket{a_0} \,
  \stackrel{t}{\longrightarrow} \, \sum_n c_n 
  \ket{s_n}\ket{a_n}. 
\end{equation}
Evidently, measurement in this sense amounts to the creation of
quantum correlations between the system and the apparatus. No
individual state vector can be assigned to either the system or the
apparatus at the conclusion of the interaction. All aspects of the
problem of the quantum-to-classical transition listed above appear
here: In general the final state on the right-hand side of
Eq.~\eqref{eq:2adgdvaj} can be rewritten in different bases; it should
be possible to measure interference between different ``pointer''
positions of the apparatus; and no one outcome (or pointer position
``$n$'') has been singled out.  Inclusion of further systems, such as
a secondary apparatus or human observer, will not terminate the
resulting \emph{von Neumann chain} if these systems are treated in the
same way as before (i.e., as interacting quantum systems with globally
unitary time evolution), as already recognized by \citet{Bohr:1927:zb}
and \citet{Neumann:1932:gq} themselves.

\subsection{\label{sec:basics-decoherence}Basics of decoherence}

The origins of the decoherence program can be traced back to a paper
of \citet{Zeh:1970:yt} who emphasized that realistic macroscopic
quantum systems are never closed. If the Schr\"odinger equation is
assumed to be universally valid, Zeh then showed that such systems
must become strongly entangled with their environments, leading to a
``dynamical decoupling'' of wave-function components and the inability
of describing the evolution of the system alone by the Schr\"odinger
equation. Zeh proposed that this mechanism could help explain the
observed fragility of quantum states of macroscopic systems and the
emergence of effective superselection rules.  Zeh's work, although
supported by Wigner\footnote{In fact, Wigner abandoned his views on
  the special role of consciousness in quantum measurement
  \citep{Wigner:1961:ss} once he became aware of Zeh's ideas on
  decoherence \citep[pp. 66, 75, 215--6, 334, 338, 341, 583, 606,
  615]{Wigner:1995:jm}.}, remained in relative obscurity for the
better part of the next decade. \citet{Zeh:2006:na} recently called
this period the ``dark ages of decoherence.'' In the 1980s, crucial
progress was made through contributions by
\citet{Zurek:1981:dd,Zurek:1982:tv}, who also played an important role
in popularizing decoherence through an article published in
\emph{Physics Today} \citep{Zurek:1991:vv}.

Since then decoherence has evolved into an exponentially growing field
of research that has attracted massive attention from the
foundational, theoretical, and experimental communities. Decoherence
is a consequence of a realistic application of standard quantum
mechanics and is an experimentally well-confirmed physical process. It
is therefore neither an interpretation nor a modification of quantum
mechanics. However, the implications of decoherence are intimately
related to interpretive issues of quantum mechanics
\citep{Schlosshauer:2003:tv}, in particular to the problem of
measurement.  \citet{Bub:1997:iq} even suggested that decoherence
represents the ``new orthodoxy'' of understanding quantum mechanics,
i.e., as the working physicist's approach to motivating the postulates
of quantum mechanics from physical principles.

We shall now briefly summarize the formalism and physical mechanism of
decoherence \citep[for in-depth reviews of the field,
see][]{Joos:2003:jh,Zurek:2002:ii,Schlosshauer:2003:tv,Schlosshauer:2007:un}.
Readers familiar with decoherence may choose to skip ahead to
Sec.~\ref{sec:bohr-heis-quant}. We may define decoherence as the
practically irreversible dislocalization (in Hilbert space) of
superpositions due to ubiquitous entanglement with the
environment. The key insight is that every realistic quantum system is
open, i.e., interacts with its environment, and that such interactions
lead to entanglement between the two partners. This means that there
exists no longer a quantum state vector that could be attributed to
the system alone. In its most basic form, the process of decoherence
can be described as a von Neumann measurement of the system by its
environment,
\begin{equation}
  \label{eq:saifweiu1}
  \left( \sum_n c_n \ket{s_n} \right) \ket{E_0} \, \stackrel{t}{\longrightarrow} \,
  \ket{\Psi(t)} = \sum_n c_n \ket{s_n} \ket{E_n(t)}.
\end{equation}
The superposition initially confined to the system has now spread to
the larger, composite system--environment state. Coherence between the
components $\ket{s_1}$ can no longer be considered a property of the
system alone.  Since we have in practice rarely access to all
environmental degrees of freedom, this process is irreversible for all
practical purposes.\footnote{In fact, such effective irreversibility
  is a necessary condition for Eq.~\eqref{eq:saifweiu1} to count as a
  real, measurement-like decoherence process---otherwise it would be
  merely a case of ``virtual'' decoherence \citep[see,
  e.g.,][Sect.~2.13]{Schlosshauer:2007:un}.}  The particular basis of
the system in which the von Neumann measurement happens is determined
by the relevant Hamiltonians, typically the system--environment
interaction Hamiltonian \citep[``environment-induced
superselection'';][]{Zurek:1981:dd,Zurek:1982:tv}.

Suppose now we inquire about the consequences of the environmental
interactions for future measurements on the system. The local
measurement statistics are exhaustively contained in the reduced
density matrix \citep{Landau:1927:uy,Neumann:1932:gq,Furry:1936:pp} for
the system, obtained by tracing over the environmental degrees of
freedom,
\begin{equation}
  \label{eq:2adgdvaasklj}
\op{\rho}_\mathcal{S}(t) = \text{Tr}_\mathcal{E} \ketbra{\Psi(t)}{\Psi(t)} =
\sum_n c_m c_n^* \ketbra{s_m}{s_n} \braket{E_n(t)}{E_m(t)}.
\end{equation}
The presence of interference terms $m \not= n$ embodies the quantum
coherence between the different components $\ket{s_n}
\ket{E_n(t)}$. Concrete models for the environment show that, with
respect to the basis $\{ \ket{s_n} \}$ of the system, the
corresponding ``relative'' states \citep{Everett:1957:rw}
$\ket{E_n(t)}$ of the environment become rapidly orthogonal, meaning
that they are able to resolve the differences between the states
$\ket{s_n}$ (similar to a ``pointer'' on a scale). 

Then the reduced density matrix \eqref{eq:2adgdvaasklj} will become
rapidly orthogonal in the preferred ``pointer'' basis $\{ \ket{s_n}
\}$ dynamically selected by the system--environment interaction
Hamiltonian \citep{Zurek:1981:dd,Zurek:1982:tv},
\begin{equation}
  \label{eq:2adgdvaasklkzvjlj}
\op{\rho}_\mathcal{S}(t) \,\longrightarrow \, 
\sum_n \abs{c_n}^2 \ketbra{s_n}{s_n} .
\end{equation}
An observer of the system cannot measure interference effects in this
basis and therefore cannot empirically confirm the presence of the
superposition. However, it is important to bear in mind that in the
global system--environment state the superposition is of course still
present, or as \citet[p.~224]{Joos:1985:iu} put it, ``the interference
terms still exist, but they are not \emph{there}.'' Thus in principle
a suitable measurement (albeit in practice usually prohibitively
difficult to implement) could always confirm the existence of the
global superposition: No single outcome $\ket{s_n}$ is selected, as
required by the unitarity of the global evolution.

The environment-superselected states $\{ \ket{s_n} \}$ are also
\emph{robust} in the sense that they become least entangled with the
environment, whereas superpositions of these states are rapidly
decohered. In this way, the familiar observables such as position,
momentum, and spin are dynamically superselected through the physical
structure of the system--environment interaction. Recent related
research programs based on studies of further consequences of
environmental entanglement, such as quantum Darwinism and redundant
environmental encoding, have shown how observers can indirectly gather
information about the system without disturbing its state (see, e.g,
\citealp{Blume:2005:oo}, and references therein).

\subsection{\label{sec:emerg-class-thro}Emergence of classicality
  through decoherence}

To what extent decoherence explains the emergence of classical
structures and properties from within quantum mechanics depends to a
significant degree on the specific axiomatic and interpretive
framework of quantum mechanics that one adopts
\citep{Schlosshauer:2003:tv}. Since we are chiefly interested in the
consequences of decoherence for observers and measurements, matters
are most clear-cut if we assume the usual measurement postulates of
quantum mechanics (i.e., the projection postulate with Born's rule).
In this case the use of reduced density matrices poses no further
interpretive difficulties, and decoherence resolves the
preferred-basis problem and explains the difficulty of observing most
interference phenomena. Although the problem of outcomes is absent by
assumption, decoherence allows us to quantify when it is appropriate
to assume that an event or outcome has happened for all practical
purposes.

One may go beyond the measurement axioms and use decoherence to show
how the projection postulate may be effectively derived from the
influence of decoherence in local measurement-like interactions.
Decoherence leads to effective ensembles of quasiclassical
wave-function ``trajectories''---in phase space \citep{Paz:1993:ta} or
other bases---which can be re-identified over time and may then be
associated with observed trajectories. In this way, classical
structures and dynamics may be understood as emergent, at the level of
intrinsically local observers, from a global wave function
\citep{Zeh:1970:yt,Zeh:1973:wq,Zeh:2000:rr,Zurek:1998:re,%
  Zurek:2004:yb,Schlosshauer:2006:rt,Schlosshauer:2007:un}.  The key
interpretive difficulty concerns the question of how to reconcile the
fact that the global quantum state contains all possible trajectories
(outcomes) while only a single trajectory is observed.\footnote{To
  resolve this problem, one may have to resort to many-worlds and
  many-minds interpretations, adopt a purely epistemic interpretation
  of quantum states, introduce hidden variables, or consider
  deviations from the Schr\"odinger dynamics. See
  \citet{Schlosshauer:2003:tv,Schlosshauer:2007:un} for discussions of
  these different options in the context of decoherence.}

We shall not further concern ourselves with these interpretive
questions but rather concisely summarize, in the most
interpretation-neutral manner possible, in what sense decoherence can
be said to account for the emergence of ``classicality.'' Decoherence
leads to the dynamical superselection, \emph{within} the quantum
formalism, of certain preferred observables that correspond to the
familiar quantities of our experience (such as position and
momentum). Decoherence causes coherences between the preferred states
to become effectively uncontrollable and unobservable at the level of
the system in the sense that phase relations are dislocalized into the
composite system--environment state.\footnote{See
  \citet{Zurek:2004:yb} for an approach to formalizing this local
  inaccessibility of phase relations without resorting to reduced
  density matrices.} Decoherence thus accounts for classicality in so
far as it describes effective restrictions on the superposition
principle for subsystems interacting with other systems described by a
larger composite Hilbert space. Save for the fundamental problem of
outcomes, this arguably explains how to reconcile our experience of
robust, seemingly measurement-independent states characterized by a
small set of definite physical quantities with the fact that the
linear Hilbert-space formalism seems to theoretically admit a vast
number of never-observed quantum states.

\section{\label{sec:bohr-heis-quant}Bohr and Heisenberg on the
  quantum--classical divide}

\subsection{\label{sec:bohr-quant-class}Bohr and the
  quantum--classical divide}

Much has been made of the fact that Bohr's epistemological approach to
quantum mechanics was opposed to von Neumann's formal measurement
scheme in which these systems were treated as unitarily evolving and
interacting quantum systems, i.e., on an equal footing with
microscopic systems. For Bohr, quantum mechanics, as a universally
valid theory, certainly \emph{could} be employed to describe the
interaction of the system under investigation and the measuring
instrument, but in doing so one was precluded from treating the
measuring instrument \emph{as} a measuring instrument. To this extent
Bohr's approach was fundamentally different to that of von Neumann,
taking the ``classicality'' of the measuring instrument as something
we must assume \emph{a priori}. Hence it is often asserted that there
was an irreducible divide for Bohr between the quantum and classical
realms. According to Jammer:
\begin{quote}
  Contrary to Planck and Einstein, Bohr did not try to bridge the gap
  between classical and quantum, but from the very beginning of his
  work, he searched for a scheme of quantum conceptions which would
  form a system just as coherent, on the one side of the abyss, as
  that of classical notions on the other
  \cite[p.~86]{Jammer:1966:cd}.\\
  Later Bohr's conception of classical mechanics, which gave deeper
  significance to his previous ideas on the irreconcilable disparity
  between classical and quantum theory, precluded, now on
  epistemological grounds, the possibility to interpret the
  correspondence principle as asserting the inclusion of classical
  mechanics within quantum theory \cite[p.~117]{Jammer:1966:cd}.
\end{quote}
This passage captures what many physicists have understood to be
Bohr's point of view. But in the time since Jammer wrote this passage
in 1966, there has been a considerable effort on the part of scholars
to come to a deeper understanding about precisely what Bohr meant. In
the late 1960s Paul \cite{Feyerabend:1968:im,Feyerabend:1969:im}
complained that Bohr's views had been systematically distorted by both
his critics and his followers, more interested in pursuing their own
philosophical agendas than seriously understanding what he had to
say. With this in mind Feyerabend urged physicists and philosophers to
go ``back to Bohr.'' A concerted effort to make philosophical sense of
Bohr's philosophical writings gathered momentum after the Bohr
centennial in 1985 \citep[pp. 201--2]{Howard:1994:lm}.

In spite of the attention Bohr's writings have received over the past
three decades, scholarly opinion on how we should understand his
thinking remains divided. This is not the place to discuss the
different interpretations of Bohr's thought, suffice it to say that
there is widespread agreement now that complementarity does not fit
neatly with the views of the logical empiricists, nor should it
necessarily be characterized as anti-realist. Bohr's viewpoint,
articulated in his reply to the EPR challenge, was that it is not
simply the case that we cannot \emph{measure} the two well-defined
attributes of an object, but rather that the mutually exclusive
experimental arrangements serve to \emph{define the very conditions}
under which we can unambiguously employ such classical concepts as
position and momentum \citep{Bohr:1949:mz}. The key point for Bohr
then is that quantum mechanics reveals to us the previously
``unrecognized presuppositions for an unambiguous use of our most
simple concepts'' \citep[pp.~289--90]{Bohr:1937:km}. We cannot ascribe
to a particle a ``position'' in space or a ``momentum'' independently
of the specific experimental conditions under which we observe the
particle.

\subsection{\label{sec:bohr-isol-syst}Bohr, isolated systems, and entanglement}

It has sometimes been argued that the characteristic feature of
quantum mechanics crucial for decoherence---namely
entanglement---plays effectively no role in Bohr's thinking. However,
a closer reading of Bohr suggests this is not so. Indeed, Bohr's
understanding of the quantum--classical divide turns out to depend on
his recognition of the nonseparability between object and instrument
in the act of measurement.

Early on Bohr recognized that the classical concept of an isolated
``object'' which has a well-defined ``state'' and which interacts with
a measuring instrument is rendered problematic in quantum
mechanics. This is grounded in the fact that, as Born observed in
1926, in quantum mechanics ``one cannot, as in classical mechanics,
pick out a state of one system and determine how this is influenced by
a state of the other system since all states of both systems are
coupled in a complicated way'' \citep[pp.~52--53]{Born:1926:en}. As
Bohr noted, this paradoxical situation in quantum mechanics has
serious implications for the concept of observation. In order to
observe a quantum system we must interact with it using some device
serving as a measuring instrument. On the one hand, in order to
observe something about an electron, say its momentum, we must
\emph{assume} that the electron possesses an independent dynamical
state (momentum), which is in principle distinguishable from the state
of the instrument with which it interacts. On the other hand, such an
interaction, if treated quantum-mechanically, destroys the
separability of the object and the instrument, since the resulting
entanglement between the two partners means that they must be
described by a single composite nonseparable quantum state. Such
entangled states represent quantum correlations between the two
systems that frequently embody entirely new physical properties for
the composite system that are not present in any of the subsystems. In
some sense, the two entangled partners have thus become a single
quantum-mechanical system.

For Bohr, this lay at the heart of the epistemological paradox of
quantum mechanics. Bohr regarded the condition of isolation to be a
simple logical demand, because, without such a presupposition, an
electron cannot be an ``object'' of empirical knowledge at all. ``The
crucial point'' he explained in 1949, is that contrary to the
situation in classical physics, in quantum mechanics we are confronted
with the ``\emph{impossibility of any sharp separation between the
  behavior of atomic objects and the interaction with the measuring
  instruments which serve to define the very conditions under which
  the phenomena appear}'' (\citealp[p.~210]{Bohr:1949:mz}, emphasis in
original). Bohr had earlier emphasized this point at the 1936 ``Unity
of Science'' congress, where he had explained:
\begin{quote}
  A still further revision of the problem of observation has since
  been made necessary by the discovery of the universal quantum of
  action, which has taught us that the whole mode of description of
  classical physics \dots retains its adequacy only as long as all
  quantities of action entering into the description are large
  compared to Planck's quantum \dots This circumstance, at first sight
  paradoxical, finds its elucidation in the recognition that in this
  region [where classical mechanics breaks down] it is no longer
  possible sharply to distinguish between the autonomous behavior of a
  physical object and its inevitable interaction with other bodies
  serving as measuring instruments, the direct consideration of which
  is excluded by the very nature of the concept of observation itself
  \citep[p.~290]{Bohr:1937:km}.
\end{quote}
The ``whole mode of description of classical physics'' to which Bohr
refers in this passage is nothing other than the condition of
separability. It is not merely that the act of measurement influences
or disturbs the object of observation, but that it is no longer
possible to distinguish between the object and its interaction with
the device serving as a measuring instrument. A quantum-mechanical
treatment of the observational interaction would paradoxically make
the very distinction between object and instrument ambiguous. However,
such a distinction is a necessary condition for empirical
inquiry. After all, an experiment is carried out precisely to reveal
information about some atomic object. As Bohr was to put it, only so
far as we can neglect the ``interaction between the object and the
measuring instrument, which unavoidably accompanies the establishment
of any such connection'' can we ``speak of an autonomous space-time
behavior of the object under observation''
\citep[p.~291]{Bohr:1937:km}. To speak of an interaction between two
separate systems---an object and measuring instrument---is to speak in
terms of classical physics.

\subsection{\label{sec:heisenberg-cut-bohr}The Heisenberg cut and the
  Bohr--Heisenberg disagreement about its ``shiftiness''}

Bohr's view made an immediate and lasting impression on Heisenberg. As
Heisenberg observed, it follows then that from an epistemological
point of view ``a peculiar schism in our investigations of atomic
processes is inevitable'' \citep[p.~15]{Heisenberg:1952:mn}. In the
discussions at the Como conference in September 1927, Heisenberg
explained that in ``quantum mechanics, as Professor Bohr has
displayed, observation plays a quite peculiar role.'' In order to
observe a quantum-mechanical object, ``one must therefore cut out a
partial system somewhere from the world, and one must make
`statements' or `observations' just about this partial system''
\citep[p.~141]{Bohr:1985:mn}. The existence of ``the cut between the
observed system on the one hand and the observer and his apparatus on
the other hand'' is a necessary condition for the possibility of
empirical knowledge. Without the assumption of such a divide we could
not speak of the ``object'' of empirical knowledge in quantum
mechanics. Heisenberg emphasized the significance of the cut
[\emph{Schnitt}] throughout the 1930s. In his lecture on ``Questions
of Principle in Modern Physics'' delivered in November 1935 in Vienna,
Heisenberg explained:
\begin{quote}
  In this situation it follows automatically that, in a mathematical
  treatment of the process, a dividing line must be drawn between, on
  the one hand, the apparatus which we use as an aid in putting the
  question and thus, in a way, treat as part of ourselves, and on the
  other hand, the physical systems we wish to investigate. The latter
  we represent mathematically as a wave function. This function,
  according to quantum theory, consists of a differential equation
  which determines any future state from the present state of the
  function \dots The dividing line between the system to be observed
  and the measuring apparatus is immediately defined by the nature of
  the problem but it obviously \emph{signifies no discontinuity of the
    physical process}. For this reason there must, within certain
  limits, exist complete freedom in choosing the position of the
  dividing line (\citealp[p.~49]{Heisenberg:1952:mk}, emphasis added).
\end{quote}
This point had been emphasized in his lecture the previous year, in
which Heisenberg argued that ``there arises the necessity to draw a
clear dividing line in the description of atomic processes, between
the measuring apparatus of the observer which is described in
classical concepts, and the object under observation, whose behavior
is represented by a wave function''
\citep[p.~15]{Heisenberg:1952:mn}. This was the central theme in an
unpublished paper, written in 1935 entitled \emph{Ist eine
  deterministische Erg\"anzung der Quantenmechanik m\"oglich?}, in
which Heisenberg outlined his own response to the criticisms of
quantum mechanics which had emerged from such physicists as von Laue,
Schr\"odinger and Einstein in the 1930s. A draft of the paper is
contained in his letter to Pauli on 2 July 1935 \citep[pp.~409--18
{[}item 414{]}]{Pauli:1985:za}. The paper, which was written at
Pauli's urging, argued that a deterministic completion of quantum
mechanics is in principle impossible. This is because in quantum
mechanics we must draw a cut between the quantum-mechanical system to
be investigated, represented by a wave function in configuration
space, and the measuring instrument described by means of classical
concepts. The critical point for Heisenberg is that ``this cut can be
shifted arbitrarily far in the direction of the observer in the region
that can otherwise be described according to the laws of classical
physics,'' but of course, ``the cut cannot be shifted arbitrarily in
the direction of the atomic system''
\citep[p.~414]{Heisenberg:1985:zq}. No matter where we chose to place
the cut, classical physics remains valid on the side of the measuring
device, and quantum mechanics remains valid on the side of the atomic
system.

There were, however, key differences, which have often been
overlooked, between Bohr and Heisenberg in their respective analyses
of the quantum--classical divide. The crucial point seems to have been
that for Heisenberg the location of the cut ``cannot be established
physically''---it represents no physical discontinuity---``and
moreover it is precisely the arbitrariness in the choice of the
location of the cut that is decisive for the application of quantum
mechanics'' \citep[p.~416]{Heisenberg:1985:zq}. We know from an
exchange of correspondence in the 1930s that Bohr objected to
Heisenberg's view that the cut could be shifted arbitrarily far in the
direction of the apparatus (\citealp{AHQP:1986:op}: Heisenberg to Bohr
10 August 1935, Bohr to Heisenberg 10 September 1935, Bohr to
Heisenberg 15 September 1935, Heisenberg to Bohr 29 September
1935). Heisenberg explained to Bohr that without such a presupposition
one would have to conclude that there exist ``two categories of
physical systems---classical and quantum-mechanical ones''
\citep[Heisenberg to Bohr 29 September 1935]{AHQP:1986:op}. Heisenberg
acknowledged that strictly speaking, however, the laws of quantum
mechanics are applicable to \emph{all} systems (including the
measuring instrument). As Heisenberg pointed out in the discussions
that followed Bohr's Como paper in September 1927: ``One may treat the
whole world as \emph{one} mechanical system, but then only a
mathematical problem remains while access to observation is closed
off'' \citep[p.~141]{Bohr:1985:mn}.\footnote{Interestingly, Everett's
  relative-state interpretation \citep{Everett:1957:rw}---and its
  subsequent development into many-worlds
  \citep{DeWitt:1970:pl,DeWitt:1971:pz,Deutsch:1985:rx} and many-minds
  interpretations \citep{Lockwood:1996:pu,Zeh:2000:rr})---have taken
  precisely the route of considering a single closed
  quantum-mechanical system, containing, among other things, the
  observers themselves, and then try to account for observers and
  their observations from within this formalism. Heisenberg's charge
  that in this case ``only a mathematical problem remains'' would then
  be transformed into a merit by Everett's idea of letting the
  quantum-mechanical formalism provide its own interpretation---a
  program that Bohr would have likely disagreed with, since for him
  the formalism by itself would become meaningless without the prior
  assumption of classical concepts.}  It was therefore an
\emph{epistemological} condition that one had to introduce the cut
into the quantum-mechanical description.

The view outlined by Heisenberg leaves it ambiguous under what
circumstances we are entitled to consider the apparatus
``classically,'' and under what circumstances it should be treated as
a ``quantum-mechanical'' system. Certain physicists were also troubled
by how we are to understand the interaction between a
``quantum-mechanical'' system and a ``classical'' measuring apparatus.
Such an interaction does not appear to be subsumed under either
classical or quantum theory. Moreover the division between the quantum
and the classical realms seems to coincide exactly with the
object--instrument and microscopic--macroscopic distinctions. It
appears odd that there should be a fundamental difference between
microscopic and macroscopic systems.

Heisenberg's treatment of the cut between the ``quantum'' object and
the ``classical'' measuring device has often been taken as a faithful
representation of Bohr's most carefully considered view of the problem
of measurement. However, Heisenberg frequently alluded to his
disagreement with Bohr on this matter in the 1930s. In \emph{Physics
  and Philosophy}, he explained that ``Bohr has emphasized that it is
more realistic to state that the division into the object and rest of
the world is not arbitrary'' and the object is determined by the very
nature of the experiment \citep[p.~24]{Heisenberg:1989:zb}. Writing to
Heelan in 1975, Heisenberg explained that he and Bohr had never really
resolved their disagreement about ``whether the cut between that part
of the experiment which should be described in classical terms and the
other quantum-theoretical part had a well defined position or not.''
In his letter Heisenberg stated that he had ``argued that a cut could
be moved around to some extent while Bohr preferred to think that the
position is uniquely defined in every experiment''
\citep[p.~137]{Heelan:1975:kk}. Weizs\"acker also later recalled that
Heisenberg had disagreed with Bohr over the cut in the 1930s
\citep[p.~283]{Weizsacker:1987:ym}. The Heisenberg--Bohr exchange
would seem to suggest that for Bohr, the quantum--classical
distinction corresponds to something ``objective,'' and is not merely
an arbitrary division.

\subsection{\label{sec:howards-reconstr-boh}Howard's reconstruction of
  Bohr's doctrine of classical concepts and improper mixtures in
  decoherence}

So how are we to understand Bohr's own view of the cut argument, and
what does this have to do with the modern decoherence approach? An
important clue can be found in a paper in which \citet{Howard:1994:lm}
suggested a new interpretation (termed a ``reconstruction'') of Bohr's
classical concepts. In particular, Howard addressed the question of
what Bohr meant when he insisted on the use of a classical
description, and the question of precisely where such a classical
description is to be employed. He argued that a careful reconstruction
of Bohr's views shows that for him ``the classical/quantum
distinction'' did not exactly coincide with the ``instrument/object
distinction'' whereas this seems to have been precisely the case for
Heisenberg. Bohr, it seems, was keen to avoid the mistaken impression
that the ``classical'' instrument somehow interacts with the
``quantum'' object. According to Howard,
\begin{quote}
  it is widely assumed that Bohr's intention was that a classical
  description be given to the measuring apparatus in its entirety, a
  quantum description being given presumably, to the observed object
  in its entirety. On this view, the classical/quantum distinction
  would coincide with the instrument/object distinction; hence, its
  designation in what follows as the ``coincidence interpretation'' of
  the doctrine of classical concepts. I will argue instead that the
  two distinctions cut across one another, that Bohr required a
  classical description of \emph{some}, but not necessarily
  \emph{all}, features of the instrument and more surprisingly,
  perhaps, a classical description of some features of the observed
  object as well. More specifically I will argue that Bohr demanded a
  classical description only of those properties of the measuring
  instrument that are correlated, in the measurement interaction, with
  the properties of the observed object that we seek to measure; and
  that this implies, as well, a classical description of the
  associated measured properties of the observed object itself
  \citep[p.~203]{Howard:1994:lm}.
\end{quote}
Howard's reconstruction provides a new reading of many crucial
passages from Bohr's writings, in which ``the classical/quantum
distinction corresponds to an objective feature of the world.''
Furthermore, this reading of Bohr makes it clear that there is no
explicit or implicit appeal to the vague notion that the measuring
instrument is a ``macroscopic'' object, having certain dimensions
\citep[p.~211]{Howard:1994:lm}. Howard argues that Bohr's classical
descriptions may be interpreted---in agreement with, but not forced
out by, Bohr's own writings---as the appropriate use of proper
mixtures for the measured system in place of the global pure state
describing the entangled system--apparatus state which is produced as
the result of a von Neumann measurement. Accordingly, this transition
is motivated by Bohr's insistence on an unambiguous and objective
description of quantum phenomena, which requires the classical concept
of separability between the observed (the system) and the observer
(the apparatus). However, the feature of quantum entanglement
shows---as evidenced by examples such as EPR \citep{Einstein:1935:dr}
and von Neumann's measurement scheme with its famous application to
Schr\"odinger's cat \citep{Schrodinger:1935:gs}---that such
separability no longer holds. As discussed in
Sec.~\ref{sec:probl-quant-class}, in a von Neumann measurement,
\begin{equation}
\label{eq:dfkasjg2}
\ket{\Psi_0} = \left( \sum_n c_n \ket{s_n} \right)
\ket{a_0} \longrightarrow \ket{\Psi} =\sum_n c_n \ket{s_n} \ket{a_n}, 
\end{equation}
the final joint system--apparatus state does not factor into a
tensor-product state $\ket{s'}\ket{a'}$, and thus no individual
quantum state can be attributed to either the system or the apparatus.
Evidently, the classical concept of independence and separability
between the system and the apparatus no longer holds.

Howard suggests that Bohr's classical concepts may be identified with
the selection of \emph{subensembles that are appropriate to the
  measurement context} (i.e., in Bohr's terminology, that are
appropriate to the particular ``experimental arrangement''), in the
following sense. Consider the density matrix corresponding to the
final state on the right-hand side of Eq.~\eqref{eq:dfkasjg2},
\begin{equation}
\label{eq:dfkasjgrs}
\op{\rho}_\mathcal{SA} = \ketbra{\Psi}{\Psi} = \sum_{nm} c_nc_{m}^*
\ketbra{s_n}{s_{m}}\otimes \ketbra{a_n}{a_{m}} \not= 
\op{\rho}_\mathcal{S} \otimes \op{\rho}_\mathcal{A},
\end{equation}
and suppose that the measurement of interest is described by the
apparatus observable $\op{O} = \sum_m o_m \ketbra{a_m}{a_m}$. Then the
statistics of such a measurement will be exhaustively described by the
subensemble
\begin{equation}
\label{eq:dfkasjgrsjghs}
\op{\rho}_{\mathcal{SA}\vert_{\op{O}}} = \sum_{n} \abs{c_n}^2
\ketbra{s_n}{s_{n}} \otimes \ketbra{a_n}{a_{n}} \equiv \sum_{n}
\abs{c_n}^2 \op{\rho}_{\mathcal{SA}\vert_{\op{O}}}^{(n)}. 
\end{equation}
Here the interference terms $n \not= m$ have been \emph{a priori}
neglected, since they cannot be measured by the chosen observable
$\op{O}$. For all purposes of this particular measurement, an
\emph{ignorance interpretation}
\citep{Espagnat:1966:mf,Espagnat:1988:cf,Espagnat:1995:ma} is then
attached to the conditional density matrix
\eqref{eq:dfkasjgrsjghs}. That is,
$\op{\rho}_{\mathcal{SA}\vert_{\op{O}}}$ is interpreted as describing
a situation in which the system--apparatus combination is in
\emph{one} of the pure states $\ket{s_n} \ket{a_n}$ but we simply do
not know in which. In this way, the density matrix
\eqref{eq:dfkasjgrsjghs} is by assumption (associated with Bohr's
assumption of ``classical concepts'') taken to represent a
\emph{proper} (classical) mixture of the ``outcome states'' $\ket{s_n}
\ket{a_n}$ for a measurement (``properization'' of the full density
matrix conditioned on the particular measurement).

According to Howard, Bohr's mutually exclusive experimental
arrangements may then be identified with the choice of such effective
mixtures conditioned on the particular measurement setup. Thus we may
say that the different decompositions of the global density matrix
into proper subensembles correspond to the different observables
measured by each of the arrangements. No single ``properized'' mixture
of the form \eqref{eq:dfkasjgrsjghs} will give the correct statistics
for all possible observables but will suffice for the measurement of
observables codiagonal in the basis used in expanding the
mixture. This ties in with Bohr's notion of the existence of classical
measurement apparatuses: In Howard's reading, the assumption of
classicality for such apparatuses would then correspond to the
existence of preferred apparatus observables whose eigenbases
determine the particular ``properized'' mixtures, which in turn
exhaustively describe the statistics of these particular
measurements.

Howard's reconstruction actually fits rather nicely with what we know
of the disagreement between Bohr and Heisenberg concerning the cut
between ``object'' and the instrument.  It also allows us to draw a
particularly interesting connection to decoherence, and in fact gives
Bohr's ``classical concepts'' (interpreted in the sense of Howard's
reconstruction) a more precise meaning. In Howard's picture, for Bohr
the choice between different ``properized'' mixtures was simply a
result of knowing which observable was measured by a particular
experimental arrangement. However, first, what determines this
observable on physical grounds? And second, in many cases the
formalism allows us to rewrite such a mixture in many different bases
(e.g., for Bell states) and thus does not uniquely fix the basis which
supposedly should correspond to a particular experimental
arrangement. How can one circumvent this problem of basis ambiguity as
posed by the formalism of quantum mechanics? Third, what precisely
justifies neglecting the interference terms in the global density
matrix?

To all questions decoherence provides an answer. In any realistic
account of measurement, we ought to include further interactions with
the environment. As a consequence of decoherence, there will be at
least one preferred basis in which the interference terms between
different one-to-one quantum-correlated system--apparatus states in
the reduced system--apparatus density matrix will be sufficiently
small in order to be neglected in practice.  We thus arrive at a
system--apparatus density matrix that is formally identical to
\eqref{eq:dfkasjgrsjghs}. The relevant observable is determined by the
structure of the apparatus--environment interaction Hamiltonian
\citep{Zurek:1981:dd,Zurek:1982:tv}, while in other bases
system--apparatus correlations would be rapidly destroyed by the
environment and therefore the apparatus could not function reliably
\citep[this is the ``stability criterion'' introduced
by][]{Zurek:1982:tv}. The preferred (environment-superselected) basis
may then be regarded as corresponding to Bohr's notion of a particular
physical arrangement that is used to measure the system. This allows
us to explain why measurement devices appear to be designed to measure
certain physical quantities but not others, while in the absence of
decoherence we would in general face the preferred-basis problem
\citep{Zurek:1982:tv,Schlosshauer:2003:tv}. Decoherence thus supplies
a physical criterion for the choice of the particular ``properized''
ensemble in Howard's reconstruction.

This dynamical picture also allows us to precisely quantify the
location of the Heisenberg cut, or, in the terminology of Howard's
reconstruction, of \emph{when} and \emph{where} we may make the
replacement of the density matrix by a ``properized'' mixture. In any
given physical situation, we can (at least in principle) model the
relevant interactions between the subsystems and thus, for each chosen
subsystem, precisely determine the degree to which certain
interference effects may be observable in a local measurement
performed on this subsystem. This allows us to quantify the degree to
which the subsystem is rendered effectively classical with respect to
different local observables. This ability to precisely pinpoint the
location of the quantum--classical divide by taking into account the
relevant decoherence effects represents an enormous progress over the
Heisenberg picture, where it was simply left to the judgment of the
observer where to place the cut.

One important point, however, remains. Reduced density matrices
obtained by tracing out the environment are not
ignorance-interpretable since, as Pessoa Jr.\
\citeyearpar[p.~432]{Pessoa:1998:yl} put it, ``taking a partial trace
amounts to the statistical version of the projection postulate.'' By
contrast, the system--apparatus density matrices in the Bohr--Howard
picture are derived from simply neglecting the interference terms and
then assuming that the resulting mixture is
ignorance-interpretable. This involves a conceptual leap just as
severe as that of choosing to interpret reduced density matrix as
ignorance interpretable. In both cases, quantum mechanics forbids us
to attach such an interpretation. Scholars working on decoherence have
been (at least lately) very careful in making this point.  Howard
emphasizes this issue, too, but his reconstruction is precisely all
about associating Bohr's assumption of classical concepts with a
deliberately ignorance of this point.

We see that Howard's reconstruction has the merit of providing a
specific formalization of Bohr's notion of classical concepts, and
decoherence shows how this formalization can be physically motivated,
justified, and quantified. If we follow Howard and take Bohr's
classical concepts as amounting to a replacement of quantum-mechanical
ensembles by classical mixtures that depend on the measurement
context, then decoherence indeed allows us to derive these concepts
(modulo the fundamental problem of the transition to
ignorance-interpretable mixtures). Or, put differently, decoherence
allows us to tell a dynamical, physical story of these concepts.

\subsection{\label{sec:heis-entangl-extern}Heisenberg, entanglement,
  and the external world}

Looking back over the history of the foundations of quantum mechanics,
we can now see the crucial obstacle to an understanding of the
quantum-to-classical transition was the erroneous assumption that we
can treat quantum systems as isolated from the environment. It is true
that Bohr had earlier understood very well that the properties
exhibited by quantum systems cannot be separated from the experimental
conditions under which they are observed. But he does not appear to
have extended this argument to the measuring apparatus and its
environment: Bohr was simply content to assume the classicality of the
experimental apparatus. The recognition that it is precisely the
openness of quantum systems and the resulting environmental
entanglement that may explain how these systems become effectively
classical was the crucial insight in the decoherence account. 

However, there do appear to be anticipations of the relevance of the
environment in certain passages in Heisenberg's writings from the
1950s. In his contribution to the volume commemorating Bohr's
seventieth birthday, Heisenberg gave his most systematic defense of
what he referred to as the ``Copenhagen interpretation of quantum
mechanics.'' In the paper Heisenberg remarked that many authors had
pointed out that the so-called ``reduction of wave-packets cannot be
deduced from the Schr\"odinger's equation'' and that to this extent a
number of physicists had drawn the conclusion ``that there is an
inconsistency in the `orthodox' interpretation''
\citep[p.~23]{Heisenberg:1955:lm}. Yet, as Heisenberg emphasized, the
reason that the quantum-mechanical treatment of the ``interaction of
the system with the measuring apparatus'' does not of itself ``as a
rule lead to a definite result (e.g. the blackening of a photographic
plate)'' is that in such a treatment ``the apparatus and the system
are regarded as cut off from \emph{the rest of the world} and treated
as a whole according to quantum mechanics''
\citep[p.~22]{Heisenberg:1955:lm}. However, as Heisenberg explained:
``If the measuring device would be isolated from the rest of the
world, it would be neither a measuring device nor could it be
described in the terms of classical physics at all''
\citep[p.~24]{Heisenberg:1989:zb}. Somewhat surprisingly Heisenberg
attributed this view to Bohr:
\begin{quote}
  Bohr has rightly pointed out on many occasions that the connection
  with the external world is one of the necessary conditions for the
  measuring apparatus to perform its function, since the behavior of
  the measuring apparatus must be capable of being \dots described in
  terms of simple [classical] concepts, if the apparatus is to be used
  as a measuring instrument at all. And the \emph{connection with the
    external world} is therefore necessary \dots We see that a system
  cut off from the external world \dots cannot be described in terms
  of classical concepts \citep[pp.~26--7, emphasis
  added]{Heisenberg:1955:lm}.
\end{quote}
Here Heisenberg appears to connect the fact that the measuring
instrument cannot be isolated from the rest of the world with the need
to use classical concepts in the description of the experiment. In
\emph{Physics and Philosophy} Heisenberg makes more explicit this
connection between the inseparability of the ``system'' and the
``external world'' and the quantum-to-classical transition:
\begin{quote}
  Again the obvious starting point for the physical interpretation of
  the formalism seems to be the fact that mathematical scheme of
  quantum mechanics approaches that of classical mechanics in
  dimensions which are large compared to the size of atoms. But even
  this statement must be made with some reservations. Even in large
  dimensions there are many solutions of the quantum-mechanical
  equations to which no analogous solutions can be found in classical
  physics. In these solutions the phenomenon of the ``interference of
  probabilities'' would show up \dots [which] does not exist in
  classical physics. Therefore, even in the limit of large dimensions
  the correlation between the mathematical symbols, the measurements,
  and the ordinary concepts [i.e., the quantum-to-classical
  transition] is by no means trivial. In order to get at such an
  unambiguous correlation one must take another feature of the problem
  into account. It must be observed that the system which is treated
  by the methods of quantum mechanics is in fact a part of a much
  bigger system (eventually the whole world); it is interacting with
  this bigger system; and one must add that the microscopic properties
  of the bigger system are (at least to a large extent) unknown. This
  statement is undoubtedly a correct description of the actual
  situation \dots The interaction with the bigger system with its
  undefined microscopic properties then introduces a new statistical
  element into the description \dots of the system under
  consideration. In the limiting case of the large dimensions this
  statistical element destroys the effects of the ``interference of
  probabilities'' in such a manner that the quantum-mechanical scheme
  really approaches the classical one in the limit
  \citep[pp.~121--2]{Heisenberg:1989:zb}.
\end{quote}
In this intriguing passage Heisenberg recognizes that it cannot simply
be the macroscopic dimensions of the measuring apparatus that ensure
that the apparatus becomes effectively classical. In the
reconstruction of Bohr's view of the quantum--classical divide
presented earlier, attention is restricted to the interaction between
the system and the apparatus. There the description of the measured
system by a proper mixture is justified by referring to the need for
an objective account of experimental results. In the passages quoted
above, however, Heisenberg enlarges the system--apparatus composite to
include couplings to further degrees of freedom in the environment
(the ``external world''). This is a most interesting point. Of course,
for practical purposes, Heisenberg admits that we often treat the
quantum system and the measuring apparatus as isolated from the rest
of the world. But the quantum-to-classical transition depends on ``the
underlying assumption,'' implicit in the Copenhagen interpretation,
``that the interference terms are in the actual experiment removed by
the partly undefined interactions of the measuring apparatus, with the
system and with \emph{the rest of the world} (in the formalism, the
interaction produces a `mixture')'' \citep[p.~23]{Heisenberg:1955:lm}.

Heisenberg's account may be read as anticipating results of the
decoherence program in two ways. First, in the formal description of
decoherence, the environment is traced out, arriving at an improper
mixtures, which is then, for all practical purposes of statistical
prediction, interpreted as a proper mixture. Second, because the
Quantum Darwinism program \citep{Blume:2005:oo} precisely shows how
the environment plays the role of an information channel that
``objectifies'' (in an effective sense) the information represented by
the system--apparatus quantum correlations through redundant encoding.
The account also shows some interesting parallels to the distinction
between decoherence and classical noise. Noise describes a situation
in which the system is perturbed by the environment. However,
decoherence corresponds to a measurement-like process in which the
system perturbs the environment, in the sense that the superposition
initially confined to the system spreads to the system--environment
combination. The nonlocal nature of quantum states then implies that
this ``distortion'' of the environment by the system in turn
influences the observable properties at the level of the system (as
formally described by the reduced density matrix).

However, while Heisenberg emphasizes the importance of environmental
interactions, nowhere does he explicate the role of entanglement
between the system and the environment as the crucial point in the
emergence of classicality in the system. In discussing the extent to
which quantum mechanics gives an ``objective'' description of the
world, Heisenberg draws attention to the fact that ``classical physics
is just that idealization in which we speak about the parts of the
world without any reference to ourselves''
\citep[p.~22]{Heisenberg:1989:zb}. Here Heisenberg is careful to avoid
the impression that quantum mechanics contains any ``genuine
subjective features''---he categorically denies that the mind of the
observer plays any crucial role in the measurement process. But he
does suggest that ``quantum theory only corresponds to this ideal [of
the separability of `objects'] as far as possible.''  Heisenberg
maintains that the somewhat arbitrary ``division of the world into the
`object' and the rest of the world'' is the starting point of quantum
mechanics. To this extent, he seems to have been of the view that such
a division between ``object'' and the ``rest of the world'' was
indispensable for physics, \emph{in spite of} the fact that quantum
systems exhibit radical nonseparability
\citep[p.~23]{Heisenberg:1989:zb}. It is the hallmark of decoherence
that it begins from the assumption that the quantum system and
apparatus cannot be isolated from the surrounding environment, and
moreover that it is precisely this feature of quantum mechanics which
results in the emergence of classicality.

Physicists such as Bohr, Heisenberg, and Schr\"odinger had recognized
the nonseparability of quantum systems---i.e., entanglement---as a
characteristic feature of quantum mechanics. For example,
Schr\"odinger, who had coined the term ``entanglement''
(\emph{Verschr\"ankung} in German) in 1935
\citep{Schrodinger:1935:gs,Schrodinger:1935:jn,Schrodinger:1936:jn},
referred to this nonseparability not as ``\emph{one} but rather
\emph{the} characteristic trait of quantum mechanics, the one that
enforces its entire departure from classical lines of thought''
\citep[p.~555, emphasis in the original]{Schrodinger:1935:jn}.  But
the feeling prevailed that entanglement was something unusual and a
peculiarly microscopic phenomenon that would have to be carefully
created in the laboratory (such as in an EPR-type experiment).
Entanglement was regarded as an essential quantum feature that would
necessarily have to be irreconcilable with classicality. These
long-held beliefs likely contributed to the comparably late
``discovery'' of decoherence \citep{Schlosshauer:2007:un}. It is
indeed a particular irony that entanglement would turn out to be not
something that had to be tamed in some way to ensure classicality but
would instead assume a key role in the emergence of classicality.

\section{\label{sec:doctr-class-conc}The doctrine of classical concepts}

\subsection{\label{sec:doctr-class-conc-revis}The doctrine of
 classical concepts revisited}

It remains here to comment on the extent to which we can say that the
decoherence account of measurement runs counter to Bohr's original
view of classical concepts. In order to do this, it seems necessary to
get a clearer picture of precisely why Bohr thought we must use
classical concepts. This is a question that has puzzled many
physicists and philosophers, but unfortunately Bohr never really
clarified his views on this issue. Howard's reconstruction of Bohr's
doctrine of classical concepts is instructive, but in the end it
simply begs the question: Why are we are entitled to replace the
quantum-mechanical density matrix with a particular ``classical''
mixture? Or, within the decoherence-based account of Howard's
reconstruction, how can we justify the use of improper mixtures
without presuming the usual axioms of measurement, which underlie the
formalism and interpretation of improper mixtures as the formal
entities that completely encapsulate all local \emph{measurement}
statistics \citep{Schlosshauer:2003:tv}?

Indeed much of what we know of Bohr's position comes from scattered
remarks, often interpreted through his contemporaries, many of whom
took up the role of Bohr's self-appointed spokesmen. The situation
becomes more difficult when we realize that Bohr's views were
appropriated by a number of different philosophical schools of thought
such as positivism, Kantianism, critical realism, linguistic idealism,
dialectical materialism, and even pragmatism. One reason that Bohr's
writings were so readily adapted to different philosophical positions
is that many of his contemporaries saw it as their task to clarify
Bohr's views, which were often not expressed quite as clearly as they
might have been. To this extent, many different versions of Bohr's
views had emerged by the 1960s, some of which were diametrically
opposed to one another.

In spite of the difficulty of this situation, we can make sense of the
different versions of Bohr's doctrine of classical concepts by
carefully drawing two useful distinctions. The first is what might be
termed the pragmatic versus categorical version of the doctrine of
classical concepts, and the second is the distinction between an
epistemological and physical formulation of the doctrine. We will
address both in turn here.

\subsection{\label{sec:pragm-vers-categ}Pragmatic versus categorical
  formulations}

In probing the doctrine of classical concepts, it is important to note
something which \citet[pp.~197--9]{Beller:1999:za} has highlighted in
her recent work. Bohr's edict that we \emph{must} use classical
concepts was sometimes given a more pragmatic interpretation by Bohr
and his followers, namely, that we simply \emph{do} use classical
concepts in describing the results of measurements in quantum
mechanics, and that this is the situation we find ourselves in. One
certainly finds this ``weaker'' version of the doctrine of classical
concepts in the writings of Rosenfeld, Heisenberg, and
Weizs\"acker. Thus the doctrine is transformed from a categorical
imperative to a pragmatic statement of the fact. In his book on the
\emph{Worldview of Modern Physics}, Weizs\"acker makes explicit this
subtle, but important, shift of emphasis:
\begin{quote}
  We ought not to say, ``Every experiment that is even possible
  \emph{must} be classically described,'' but ``Every actual
  experiment known to us \emph{is} classically described, and we do
  not know how to proceed otherwise.'' This statement is not
  sufficient to prove that the proposition is \emph{a priori} true for
  all, merely possible future knowledge; not is this demanded by the
  concrete scientific situation. It is enough for us to know that it
  is a priori valid for quantum mechanics \dots We have resolved not
  to say, ``Every experiment \emph{must} be classically described''
  but simply, ``Every experiment \emph{is} classically described.''
  Thus the factual, we might almost say historical situation of
  physics is made basic to our propositions \citep[pp.~128,
  130]{Weizsacker:1952:ym}.
\end{quote}
Here Weizs\"acker is clear that the doctrine of classical concepts has
``not logical but factual necessity.'' In other words, it is simply
the case that quantum mechanics is founded upon experimental results
that \emph{are} described by means of classical concepts. Physicists
defending Bohr's doctrine of classical concepts often resorted to this
pragmatic formulation when challenged. Heisenberg provides another
case in point. Responding to the suggestion that it might be possible
to ``depart from classical concepts'' in providing a genuinely
\emph{quantum-mechanical description} of experiments, Heisenberg
explained that classical concepts
\begin{quote}
  are an essential part of the language which forms the basis of all
  natural science. Our actual situation in science is such that we
  \emph{do} use classical concepts for the description of experiments,
  and it was the problem of quantum theory to find a theoretical
  interpretation of the experiments on this basis
  \citep[p.~23, emphasis added]{Heisenberg:1989:zb}.
\end{quote}
Here the historical and pragmatic dependence of quantum mechanics on
classical concepts is once again emphasized.

Of course simply appealing to the fact that we happen to use classical
concepts does not of itself prove that we cannot conceptualize the
world of experience in any other way. Weizs\"acker and Heisenberg
certainly recognized this, but the point they stressed was that our
current theory of quantum mechanics only corresponds to what can be
observed if we assume that the results of measurements are described
in terms of the basic concepts of classical physics. For Bohr, on the
other hand, the doctrine of classical concepts was expressed more
categorically: ``It lies in the nature of physical observation \dots
that all experience \emph{must} ultimately be expressed in terms of
classical concepts'' \citep[p.~94, emphasis added]{Bohr:1987:aw};
``the unambiguous interpretation of any measurement \emph{must} be
essentially framed in terms of classical physical theories, and we may
say that in the sense the language of Newton and Maxwell will remain
the language of physics for all time'' \citep[p.~692, emphasis
added]{Bohr:1931:ii}. This brings us back to the question posed at the
beginning of this section: Why must we interpret observations
classically? In a letter to Bohr written in October 1935,
Schr\"odinger asked Bohr why this remained one of his deepest
convictions \citep[pp.~508--9]{Bohr:1996:mn}. Bohr's reply
unfortunately does not shed much light on the matter, as he simply
restated what he saw as ``the seemingly obvious fact that the
functioning of the measuring apparatus must be described in space and
time'' \citep[pp.~511--2]{Bohr:1996:mn}.

The task of answering the question was left largely to Bohr's
followers such as Heisenberg, Weizs\"acker, Rosenfeld, and Petersen,
many of whom were in essential disagreement about the finer points of
Bohr's interpretation. Yet, in spite of the diversity of viewpoints, we
can distinguish two fundamentally different approaches to answering
the question left by Bohr. These may be labeled as the
\emph{epistemological} and \emph{physical} formulations of the
doctrine of classical concepts. A deeper understanding of these two
approaches provides an important clue to understanding what is really
meant by Bohr's doctrine of classical concepts and how it relates to
decoherence. It is to this that we now turn our attention.

\subsection{\label{sec:epist-vers-phys}Epistemological versus physical
  versions}

Much of the discussion over the extent to which decoherence marks a
break from the ``Copenhagen'' viewpoint suffers from the failure to
fully appreciate the different ways in which the question of why we
must use classical concepts in the description of experiments was
interpreted by Bohr's followers. Here we need to distinguish two
fundamental approaches.

In the first instance those who defended Bohr tended to frame the
question in \emph{epistemological} terms. This amounts to asking why
our conceptual framework is so wedded to our classical intuitions
about the world. Physicists who approached the doctrine of classical
concepts from this perspective attempted to give the doctrine a
decidedly Kantian, linguistic, or pragmatic reading. Weizs\"acker and
Heisenberg, for example, were inclined to interpret Bohr as having
``pragmatized'' or ``relativized'' Kant's philosophy
\citep{Camilleri:2005:kk}. In this context Heisenberg would pronounce
in 1934 that ``modern physics has more accurately defined the limits
of the idea of the \emph{a priori} in the exact sciences, than was
possible in the time of Kant'' \citep[p.~21]{Heisenberg:1952:mn}. Some
authors such as \citet{Faye:1991:za} have argued that Bohr himself was
influenced by Kantian tradition through his association with the
Danish philosopher H{\o}ffding.  Petersen argued that ``Bohr's
remarks'' on the indispensability of classical concepts ``are based on
his general attitude to the epistemological status of language and to
the meaning of unambiguous conceptual communication, and they should
be interpreted in that background''
\citep[p.~179]{Petersen:1968:uu}. Much has been made of Petersen's
remarks on Bohr's view of the primacy of language over reality in
quantum mechanics. Indeed on many occasions Bohr emphasized that an
``objective'' description amounts to the possibility of ``unambiguous
communication,'' and to this extent it must be expressed through
concepts of classical physics. Petersen saw Bohr's writings as having
provided a deep insight ``into the epistemological role of the
conceptual framework'' of classical physics (p.~185).

While approaching the meaning of doctrine of classical concepts from
an altogether different philosophical perspective, Rosenfeld agreed
that Bohr's primary concerns arose from ``general epistemological
considerations about the function of language as a means of
communicating experience''
\citep[p.~526]{Rosenfeld:1979:bg}. Rosenfeld argued that ``we must
make use of the concept of classical physics'' in describing
phenomena, simply because we must attempt to make ourselves understood
to other human beings, and here the concepts of classical physics
provide the means by which we can unambiguously communicate the
results of our observations. The concepts of classical physics are for
Rosenfeld not to be understood as somehow part of the \emph{a priori}
structure of the human mind, but have adapted to our experience of the
world. Given that all experimental knowledge of the atomic world
involves the amplification of effects such that they can be perceived
by human beings at the macrolevel, it should therefore come as no
surprise that the classical concepts form the basis of our description
of experience even in quantum mechanics.

Yet, the epistemological perspective invariably leads to another
question, which amounts to a reformulation of the doctrine of
classical concepts in \emph{physical} terms. We may state this as
follows: Why is the world such that the concepts of classical physics
\emph{can be} employed, at least to a very good approximation, in
certain situations?  Or, to put it another way: Why are classical
concepts applicable at all to the quantum world? This is a salient
question, given that, strictly speaking, the world \emph{is}
nonclassical. While Bohr himself did not attempt to provide such a
physical explanation for the doctrine of classical concepts, in the
1950s and 1960s a number of physicists turned their attention to
accounting for the emergence of classicality wholly within the
framework of quantum mechanics. This approach, which is closer in
spirit to the decoherence program, was pursued by those who saw
themselves as working within the ``Copenhagen'' tradition. The passage
quoted earlier from Heisenberg, in which he attempts to explain the
quantum-to-classical transition, may be taken as one example.

In the 1960s Weizs\"acker and Rosenfeld both attempted to defend this
kind of approach to the physics of the quantum-to-classical transition
as entirely in keeping with the spirit in which Bohr had intended his
doctrine of classical concepts. As Weizs\"acker put it at a colloquium
in 1968, ``the crucial point in the Copenhagen interpretation'' is
captured, ``but not very luckily expressed, in Bohr's famous statement
that all experiments are to be described in classical terms''
\citep[p.~25]{Weizsacker:1971:ll}. As a devotee of Bohr, this was a
view that Weizs\"acker endorsed wholeheartedly, but which he now
wished to justify. ``My proposed answer is that Bohr was essentially
right'' in arguing that the results of all measurements must be
classically describable, i.e. localized in space-time, ``but that he
did not know why'' \citep[p.~28]{Weizsacker:1971:ll}. The paradox at
the heart of the Copenhagen interpretation for Weizs\"acker is
therefore to be stated: ``Having thus accepted the falsity of
classical physics, taken literally, we must ask how it can be
explained as an essentially good approximation'' when describing
objects at the macrolevel. He spells this out:
\begin{quote}
  This amounts to asking \emph{what physical condition must be imposed
    on a quantum-theoretical system in order that it should show the
    features which we describe as ``classical.''} My hypothesis is
  that this is precisely the condition that it should be suitable as a
  measuring instrument. If we ask what that presupposes, a minimum
  condition seems to be that irreversible processes should take place
  in the system. For every measurement must produce a trace of what
  has happened; an event that goes completely unregistered is not a
  measurement. Irreversibility implies a description of the system in
  which some of the information that we may think of as being present
  in the system is not actually used. Hence the system is certainly
  not in a ``pure state''; we will describe it as a ``mixture.'' I am
  unable to prove mathematically that the condition of irreversibility
  would suffice to define a classical approximation, but I feel
  confident it is a necessary condition \citep[p.~29, emphasis in
  original]{Weizsacker:1971:ll}.
\end{quote}
Already in the 1960s a number of physicists had devoted themselves to
investigating the thermodynamic conditions of irreversibility that
would need to hold in order for a measurement to be registered
macroscopically as ``classical.'' In 1965 Rosenfeld conceded that ``it
is understandable that in order to exhibit more directly the link
between the physical concepts and their mathematical representation, a
more formal rendering of Bohr's argument should be attempted''
\citep[p.~536]{Rosenfeld:1979:bb}. In fact, Rosenfeld felt that this
had been carried out by \citet*{Daneri:1962:om} in their thermodynamic
analysis of the irreversible amplification process triggered by the
interaction between the quantum system and the measuring device. For
Rosenfeld this work had clarified many of the misunderstandings, which
had arisen through ``the deficiencies in von Neumann's axiomatic
treatment'' \citep[p.~537]{Rosenfeld:1979:bb}.

However, this would prove to be a controversial claim, which generated
much debate and discussion well into the 1970s.  As Jammer puts it,
``Rosenfeld's unqualified endorsement of the Daneri, Loinger, and
Prosperi measurement theory raises the question whether this is really
congenial, or at least not incompatible, with the basic tenets of the
Copenhagen interpretation'' \citep[p.~493]{Jammer:1974:pq}. Whereas
for Rosenfeld, this theory of measurement was ``in complete harmony
with Bohr's ideas'' \citep[p.~539]{Rosenfeld:1979:bb}, for Bub it
represented an approach to quantum theory fundamentally at odds with
Bohr's \citep[p. 65]{Bub:1971:ll}. While a closer historical analysis
is beyond the scope of this paper, the episode is indicative of the
attempts to reformulate Bohr's doctrine of classical concepts in the
1960s.  It was nevertheless clear to Wigner that ``the transition to a
classical description of the apparatus'' in this kind of approach was
``an arbitrary step'' which only served to ``postulate the miracle
which disturbs us'' \citep[p.~65]{Wigner:1995:jm}. It was not until
studies of decoherence were conducted that physicists were able to
give an adequate account of the elusive quantum-to-classical
transition.

\subsection{\label{sec:decoh-doctr-class}Decoherence and the doctrine
  of classical concepts}

As we have seen, decoherence provides a dynamical account of key
components of the quantum-to-classical transition, and in doing so,
clarifies many of the issues that had troubled an earlier generation
of physicists. However, the extent to which this marks a departure
from Bohr's doctrine of classical concepts depends in large part on
the way in which we interpret the doctrine. In attempting to find a
physical foundation for the doctrine, Heisenberg, Rosenfeld, and
Weizs\"acker were asking questions to which decoherence now supplies a
ready-made answer. They saw this effort as an extension of the general
line of thought initiated by Bohr, and entirely keeping with the
spirit in which he approached quantum mechanics. We may recall that
for Bohr, the experimental arrangements serve to define the very
conditions under which we can unambiguously employ such classical
concepts as position and momentum. Decoherence certainly goes beyond
anything that Bohr had to say in identifying ubiquitous and
practically irreversible entanglement with a large number of
environmental degrees of freedom as the crucial process which leads to
the emergence of classicality in quantum systems (at least in an
effective ``relative-state'' sense).  Decoherence shows the conditions
under which classicality arises in quantum mechanics, and to this
extent it may be regarded as providing a physical justification for
the \emph{pragmatic} use of classical concepts in a given experimental
situation.

Does the irreducibility of classical concepts in the quantum
description hold once we recognize that they are, so to speak,
``produced'' out of the quantum formalism, which may thus be
considered more fundamental? One may suggest that this undermines the
Copenhagen school's insistence on the epistemological primacy of
classical concepts. But since decoherence is simply a consequence of a
realistic application of the standard quantum formalism, it cannot by
itself give an interpretation or explanation of this formalism itself.
So the question remains, as it did before, whether it is possible to
render quantum mechanics as a meaningful theory about the world
without the use of classical concepts such as position and momentum
\cite[see][]{Howard:1994:lm}: In so far as decoherence depends on the
use of the quantum formalism, which must itself be given a physical
interpretation, some may suggest that the use of classical concepts
has already been presupposed. 

For example, part of the quantum formalism is usually derived through
the ``quantization'' of the classical position and momentum variables
of single particles, which then define configuration space as the
preferred arena for the wave function. In a similar manner, classical
Hamiltonians (e.g., for the harmonically bound particle) are directly
``translated'' into the operator-based quantum picture. However, as
\citet{Zeh:2003:pp} has pointed out, quantum field theory indicates
that the only fundamental quantization required is at the levels of
(postulated) spatial fields,\footnote{Somewhat misleadingly, this
  procedure is commonly referred to as the ``\emph{second}
  quantization,'' while the \emph{less} fundamental quantization of
  particle positions and momenta is denoted as the ``\emph{first}
  quantization.''} while the concept of particles can then be
\emph{derived} in terms of decoherence-induced (improper) ensembles of
narrow position-space wave packets \citep{Zeh:1993:lt}.\footnote{Early
  on, Heisenberg had contemplated the derivation of quantum mechanics
  and particle-like structures from the quantization of fields in
  three-dimensional space. This insight turned out to be particularly
  important for Heisenberg's own understanding of wave--particle
  duality (\citealp{Heisenberg:1989:zb}, pp.~86, 93--4;
  \citealp{Camilleri:2006:ym}). Heisenberg concluded that in this
  approach to quantum mechanics, the \emph{classical concept} of the
  field, not the particle, would constitute the fundamental starting
  point.}  Thus, while it is clear that at some stage of the theory we
have to identify the physical entities to which the mathematical
formalism refers, these entities will not need to take the form of
familiar classical concepts such as particles and their positions.

Therefore any assessment of the extent to which decoherence allows us
to ``derive classical concepts'' must inevitably depend on their
definition and the level at which one demands an explanation of such
concepts. Bohr and his followers in the 1930s understood this as an
\emph{epistemological}, not a physical, condition imposed on our
description of nature, while (as we have seen) later followers of Bohr
sought to give the doctrine a more pragmatic, physical
underpinning.\footnote{More recently, physicists have attempted to
  redefine Bohr's notion of complementarity in terms of entanglement
  \citep{Bertet:2001:zv}, while others now claim to derive ``the''
  Copenhagen interpretation from more fundamental principles
  \citep{Ulfbeck:2001:oo}. A Copenhagen-esque spirit may also be
  evident in the recent interpretive stance that ``quantum mechanics
  is about information'' \citep[see,
  e.g.,][]{Fuchs:2001:os,Zeilinger:2002:za,Bub:2004:lk}. For example,
  Zeilinger suggests that ``\emph{information} is the most basic
  notion of quantum mechanics, and it is \emph{information} about
  possible \emph{measurement} results that is represented in the
  quantum states. \emph{Measurement} results are nothing more than
  states of the classical \emph{apparatus} used by the
  experimentalist'' \citep[][p.~252, emphasis in the
  original]{Zeilinger:2002:za}. The epistemological constraints
  underlying Bohr's doctrine of classical concepts are in
  information-based interpretations often identified with limitations
  on the amount of information that nature if willing to proliferate,
  thus motivating the view that quantum mechanics is, at least in
  part, a theory about information. In this way, the quantum formalism
  may be \emph{reconstructed} (rather than interpreted) from
  fundamental assumptions about restrictions on the flow of
  information, expressed as information-theoretic principles
  \citep[see, e.g., the Clifton--Bub--Halvorson theorem;
  ][]{Clifton:2003:mz}. For critical assessments of such
  information-based interpretations, see, e.g.,
  \citet{Hagar:2003:mu,Hagar:2006:um,Daumer:2006:ym,Shafiee:2006:mh}.}

\section{Conclusions}

Decoherence allows us to analyze, in precise formal and quantitative
terms and wholly from within the quantum-mechanical formalism, when
and how the quantum-to-classical transition happens. It unambiguously
specifies the location of the Heisenberg cut and the conditions under
which certain superposition states---such as those of the
Schr\"odinger-cat type---can be prepared and observed, and what the
lifetimes of such states will be in a given experimental situation. It
therefore explains, for example, why superpositions of macroscopically
distinct positions of a large object are so prohibitively difficult to
prepare and maintain in practice. To our knowledge, there are no
experimental observations of quantum-to-classical processes that could
not be accounted for, at least in principle, by
decoherence.\footnote{We emphasize that this statement is independent
  of any assessment of whether and how decoherence may help solve the
  measurement problem, especially in the sense of the
  ``macro-objectification'' problem
  \citep{Jammer:1974:pq,Bassi:2000:ga,Adler:2001:us,%
    Schlosshauer:2003:tv,Zurek:2007:km}.}

Decoherence thus provides a physical, quantitative underpinning of the
quan\-tum--classical divide and the dynamics at this boundary. In
doing so, it clarifies the notion of the quantum--classical cut, which
was at the center of the disagreement between Bohr and Heisenberg in
the 1930s. As we have seen, Howard's reconstruction of Bohr's view of
the classical-quantum divide not only can help us make sense of the
historical disagreement between Bohr and Heisenberg, but also provides
an important step in reconstructing the link between Bohr's views and
decoherence. It is rather clear that the application of quantum
mechanics along the entire chain of interacting systems (Einstein's
\emph{``ganzer langer Weg,''} as recalled by
\citealp{Heisenberg:1971:mn}; see also \citealp{Zeh:2000:rr}),
including measurement devices \emph{and their environments}, turned
out to be a key point in an understanding of the quantum-to-classical
transition, ironically prominently involving the most distinctly
quantum-mechanical features of entanglement and unitary
evolution. Although Bohr did not deny that in principle such a fully
quantum-mechanical treatment was possible, he considered it
meaningless. He thereby closed himself, on grounds of a particular
philosophical stance, to the approach of subjecting the measurement
process to further quantum-mechanical analysis, \emph{in spite} of the
fact that he arguably believed quantum theory to be universal. 

As the insights brought about the study of decoherence show, Bohr's
attitude in this matter may be considered premature. It is worth
noting that Heisenberg's writings in the 1950s appear to acknowledge
that the quantum-to-classical transition can be understood only if we
bear in mind that the system comprising the quantum object and the
measuring apparatus is in reality part of a bigger system, which
ultimately encompasses the whole world. Here it seems that Heisenberg
went some way toward anticipating the results of decoherence, although
he never seems to have pursued the idea in any systematic fashion, and
he never makes explicit the importance of entanglement in the process.

If classical concepts are understood in the pragmatic sense---as
something we simply \emph{do} use---decoherence suggests why this
so. Decoherence shows that for macroscopic systems, and thus any
system that can legitimately count as a measuring instrument capable
of sufficiently amplifying measurement results as to make them
accessible to our experience, decoherence will be so strong as to
dynamically preclude most quantum states, save for those that turn out
to be precisely the (approximate) eigenstates of ``classical''
observables such as position. Furthermore, information about such
classical observables becomes amplified through redundant encoding in
the environment, thus meeting, at least in an effective and
relative-state state sense, Bohr's criteria of ``objectification'' and
``unambiguous communication'' that seem so inherently wedded to
classical physics. In this sense, decoherence indeed allows us to
formulate classical concepts in physical terms: It not only tells us
\emph{why} the concepts of classical physics are applicable in the
macroscopic situations relevant to our experience despite the
underlying quantum-mechanical description of the world, but also
\emph{when} and \emph{where} these concepts can be applied.

On the other hand, it is much more difficult to provide a reasonably
conclusive answer to the question of whether decoherence suggests that
Bohr's assumption of irreducible classical concepts as an
epistemological, metatheoretical---and thus ultimately
philosophical---construct may be redundant.  If classical concepts are
understood in Bohr's imperative sense---as something we \emph{must}
use---we still may invoke decoherence to \emph{justify} this
philosophical stance in a practical sense.  While decoherence allows
us to identify dynamically created classical structures and properties
within the quantum formalism, of course it does not---and cannot---in
and of itself provide us with an answer to the question of how to
\emph{interpret} this formalism, although it may lend additional
support to, or disqualify, certain interpretations
\citep{Schlosshauer:2003:tv}. Bohr's fundamental point was that any
interpretation of quantum mechanics must in the end fall back on the
use of classical concepts. In this sense one may suggest that
decoherence provides a physical justification for Bohr's intuition. In
fact, \citet{Zurek:2002:ii,Zurek:1998:re,Zurek:1993:pu,%
  Zurek:2004:yb,Zurek:2007:km} locates his decoherence-based
``existential interpretation'' between Bohr and Everett and relates it
to a ``neo-Copenhagen strategy.''  Such trends may further indicate
that the possibility of a peaceful coexistence between Bohr's
philosophy and decoherence could be considered more viable than it has
previously often been claimed.

\begin{ack}
  M.S.\ acknowledges support from the Australian Research Council.
\end{ack}


\end{document}